\documentclass[aps,prd,onecolumn,eqsecnum,amsmath,nofootinbib,preprintnumbers]{revtex4}%
\newcounter{example}[section]

\usepackage{ulem}
\usepackage{amsmath}
\usepackage{color,graphicx,float,subfigure}
\usepackage{amsfonts,amssymb,theorem,mathrsfs,times}
\usepackage{bm}
\usepackage{amsmath}
{\theorembodyfont{\upshape}
	}
{\theorembodyfont{\upshape}
	}
{\theorembodyfont{\upshape}
	}
{\theorembodyfont{\upshape}
	\newtheorem{dn}{Definition}}
{\theorembodyfont{\upshape}
	}
{\theorembodyfont{\upshape}
	}

\newcommand{\dalm}{\kern1pt\vbox{\hrule height 0.9pt\hbox{\vrule width
			0.9pt\hskip 2.5pt\vbox{\vskip 5.5pt}\hskip 3pt\vrule width
			0.3pt}\hrule height 0.3pt}\kern1pt}

\begin{document}
	\title{From quasi-local definitions to a dynamical potential: A unified framework for evolving circular orbits in dynamical spacetimes}
	
	%

	\author{Yong Song\footnote{e-mail
			address: syong@cdut.edu.cn}
	}

	
	\affiliation{
		College of Physics,\\
		Chengdu University of Technology, Chengdu, Sichuan 610059,
		China}
	

	\date{\today}
	
	\begin{abstract}
	The study of circular orbits is fundamental in gravitational physics, yet their definition in dynamical spacetimes remains challenging due to the lack of temporal symmetry. In this work, we establish a unified framework by commencing from the geometrically invariant quasi-local definition of a particle surface. We demonstrate that this definition naturally leads to a set of conditions that can be recast into the language of a coordinate-dependent dynamical potential. This potential serves as a practical computational tool for locating evolving circular orbits within a specific coordinate system. We rigorously prove the equivalence between the quasi-local and dynamical potential approaches in dynamical spherically symmetric spacetimes. The efficacy and self-consistency of the dynamical potential method are explicitly verified through its application to the Oppenheimer-Snyder dust collapse model, where it correctly reproduces the established evolution equations for null and timelike circular orbits. This work bridges the gap between abstract geometric definitions and concrete calculations, providing a robust and adaptable framework for analyzing orbital dynamics in time-dependent gravitational fields.
	\end{abstract}
	

	\maketitle


\section{Introduction}
Circular orbits have been a central topic in gravitational physics since the early days of general relativity. From the classic works on planetary motion in the Schwarzschild spacetime to modern research on black holes and gravitational waves, understanding circular orbits provides key insights into the nature of spacetime and gravitational interactions. In particular, in the context of black hole physics, they provide insights into the spacetime structure and geometric properties surrounding black holes. In 2019, the Event Horizon Telescope (EHT) Collaboration published the first image of the black hole at the center of the M87 galaxy~\cite{EventHorizonTelescope:2019dse}. This image clearly shows a shadowed region surrounded by a halo. The shadowed region is referred to as the black hole shadow, which is described by the photon sphere (i.e., the surface where null circular orbits are located). The halo corresponds to the black hole accretion disk, which consists of stable timelike circular orbits. Studying circular orbits is essential for understanding the physical mechanisms of accretion disks, simulating their formation, and analyzing black hole shadows~\cite{Abramowicz:2011xu,Grenzebach:2014fha}. Circular orbit studies are also vital for gravitational wave detection~\cite{Maggiore:2007ulw}, predicting gravitational wave signals from events such as black hole mergers, and inferring source parameters. Moreover, they test the strong gravitational field predictions of general relativity and explore physics beyond it~\cite{Berry:2020ntz}.

To solve for circular orbits in spacetime, two primary methods exist: the quasi-local method and the effective potential method. The quasi-local method~\cite{Claudel:2000yi,Cao:2019vlu,Kobialko:2022uzj,Qiao:2022jlu,Song:2022zns,Song:2022fdg,Bogush:2023ojz}, which examines the geometric properties of orbital surfaces, is broadly applicable to both static, stationary and dynamic spacetimes but involves complex calculations. In contrast, the effective potential method, which involves separating the geodesic equations to define the effective potential of the system~\cite{Teo:2003ltt,Cardoso:2008bp,Teo:2020sey,Perlick:2021aok} and is renowned for its simplicity and intuitiveness, is widely used in static and stationary spacetimes but has not been applicable to dynamical spacetimes until now.

In our previous work~\cite{Song:2022zns}, we provided an alternative quasi-local definition for circular orbits and demonstrated its equivalence to the definition based on the work of Ellis et al.~\cite{Claudel:2000yi,Bogush:2023ojz,Kobialko:2022uzj,Song:2022fdg}. We also found that this alternative definition is equivalent to the effective potential method in the static and stationary cases. However, extending the effective potential method to dynamical spacetimes presents significant challenges. The absence of a timelike Killing vector in dynamical spacetimes implies the non-existence of a conserved energy functional. This fundamental difference precludes a direct, symmetry-based extension of the effective potential method. Motivated by the need for a calculational framework that is both geometrically well-founded and practically applicable, we address the following questions: Can the quasi-local definition be translated into a set of conditions amenable to computation in a chosen coordinate system? If so, what is the form of these conditions, and can they be interpreted in terms of a generalized potential? In this work, we demonstrate that the answer is affirmative. We start from the quasi-local definition and systematically derive a dynamical potential formalism. This approach guarantees that the orbits computed are geometrically meaningful, while the potential itself provides the intuitive computational advantages of the traditional effective potential method. We validate this formalism in the in-going Vaidya spacetime, the Oppenheimer-Snyder (OS) collapse model and the Lema\^{i}tre-Tolman-Bondi (LTB) model, confirming its consistency with known results~\cite{Cao:2019vlu}.

This paper is organized as follows: In Section \ref{section2}, we first review the effective potential method in static spacetimes to establish concepts and notation. We then introduce the quasi-local definition of a particle surface and derive the fundamental conditions for evolving circular orbits in a dynamical spherically symmetric spacetime. In Section \ref{section3}, we show how these conditions motivate the introduction of a dynamical potential $V_{\pm}(t, r)$ and prove the equivalence between the two frameworks. Section \ref{section4} is devoted to validating the dynamical potential approach by applying it to the in-going Vaidya spacetime, the OS model and the LTB model. Section \ref{conclusion} presents our conclusions and discussions.

\section{Theoretical Framework}\label{section2}
\subsection{Effective Potential in Static Spacetimes: A Review}
We first review the effective potential method for solving circular orbits in static spherically symmetric spacetimes, as this establishes the foundation for our generalization. 

In a static spherical symmetric background, the line element in standard spherical coordinates $(t,r,\theta,\phi)$ can be written as
\begin{align}
	\label{staticmetric}
	ds^2=-f(r)dt^2+g(r)dr^2+r^2(d\theta^2+\sin^2\theta d\phi^2)\;,
\end{align}
where $f$ and $g$ are functions that depend on the radial coordinate $r$. Due to the spherical symmetry of the system, we can restrict our analysis to the equatorial plane. The Lagrangian for a test particle is then
\begin{align}
	\label{LagrangianDensity}
	\mathcal{L}=\frac{1}{2}\bigg[-f(r)\bigg(\frac{dt}{d\lambda}\bigg)^2+g(r)\bigg(\frac{dr}{d\lambda}\bigg)^2+r^2\bigg(\frac{d\phi}{d\lambda}\bigg)^2\bigg]\;,
\end{align}
where $\lambda$ is the affine parameter along the geodesic. Since the Lagrangian (\ref{LagrangianDensity}) does not depend explicitly on the variables $t$ and $\phi$, there are two conserved quantities along the geodesic
\begin{align}
	\label{Estatic}
	&p_t\equiv\frac{\partial\mathcal{L}}{\partial(\frac{dt}{d\lambda})}=-f\frac{dt}{d\lambda}=-E\;,\\
	\label{Lstatic}
	&p_\phi=\frac{\partial\mathcal{L}}{\partial(\frac{d\phi}{d\lambda})}=r^2\frac{d\phi}{d\lambda}=L\;,
\end{align}
where $E$ and $L$ are the energy and angular momentum of the particles, respectively. From the normalization of the four-velocity $(\frac{dx^\mu}{d\lambda})$, i.e., 
\begin{align}
	\label{normalization}
	g_{\mu\nu}\frac{dx^\mu}{d\lambda}\frac{dx^\nu}{d\lambda}=\epsilon\;,
\end{align}
where $\epsilon=-1, 0, 1$ correspond to the timelike, null, and spacelike geodesics, respectively, and using Eqs.(\ref{Estatic}) and (\ref{Lstatic}), we obtain
\begin{align}
	\frac{dr}{d\lambda}=\pm\sqrt{\frac{1}{g}\bigg(\frac{E^2}{f}-\frac{L^2}{r^2}+\epsilon\bigg)}\;.
\end{align}
The motion of the particles is completely governed by the effective potential, defined as
\begin{align}
	\label{V}
	V_{\pm}(r)=\pm V(r)=\pm\sqrt{\frac{1}{g}\bigg(\frac{E^2}{f}-\frac{L^2}{r^2}+\epsilon\bigg)}\;,
\end{align}
A particle moves along a circular orbit when the following two conditions are simultaneously satisfied~\cite{Jefremov:2015gza,Toshmatov:2020wky}:
\begin{itemize}
	\item[(1).]  The particle has zero radial velocity, i.e.,
	\begin{eqnarray}
		\label{V=0}
		\frac{dr_o}{d\lambda}=0\Longrightarrow V(r_o)=0\;.
	\end{eqnarray}
	\item [(2)] The particle has zero radial acceleration, i.e.
	\begin{eqnarray}
		\label{V1=0}
		\frac{d^2r_o}{d\lambda^2}\Longrightarrow \frac{dV}{d\lambda}\Longrightarrow V'(r_o)=0\;.
	\end{eqnarray}
	where $r_o$ is the radius of the circular orbit and a prime denotes differential with respect to $r_o$.
\end{itemize}
From these conditions, one can derive the equations for circular orbits in static spacetime.

Since $E$ and $L$ are conserved along the geodesic\footnote{Although $E$ and $L$ are constant along the geodesic, by considering 
	\begin{align}
		\frac{dE}{d\lambda}=0\;,\quad\mathrm{and}\quad \frac{dL}{d\lambda}=0\;,
	\end{align}
	and Eqs. (\ref{Estatic}) and (\ref{Lstatic}), one can obtain the non-trivial equations related to the $t$ and $\phi$ components of the geodesic equations.}, we have
\begin{align}
	\label{E1L1}
	\frac{dE_o}{d\lambda}=0\;,\quad\frac{dL_o}{d\lambda}=0\Longrightarrow E_o'(r_o)=L_o'(r_o)=0\;.
\end{align}
where $E_o$ and $L_o$ are the energy and angular momentum of the particle on the circular orbit. Combining Eqs.(\ref{V=0}) and (\ref{E1L1}), one can also derive the equations of the circular orbits in static spacetime. For example, from Eq.(\ref{V=0}), we obtain
\begin{align}
	E_o^2=f\bigg(\frac{L_o^2}{r^2_o}-\epsilon \bigg)\;.
\end{align}
From Eq.(\ref{E1L1}), the radius of the circular orbit satisfies
\begin{align}
	&L_o^2=\frac{\epsilon r_o^3 f'}{r_of'-2f}\;,\\
	&E_o^2=\frac{2\epsilon f^2}{r_of'-2f}\;.
\end{align}
which is consistent with the results in~\cite{Cardoso:2008bp}.

In summary, we have three equivalent sets of conditions for determining circular orbits in static spacetimes, all of which are also applicable in stationary spacetimes:
\begin{itemize}
	\item [(1).] The radial coordinate of the circular orbits does not change along the geodesic:
	\begin{align}
		\frac{dr_o}{d\lambda}=0\quad\mathrm{and}\quad \frac{d^2r_o}{d\lambda^2}=0\;.
	\end{align}
	\item [(2).] The effective potential of the circular orbits does not change along the geodesic:
	\begin{align}
		V(r_o)=0\quad\mathrm{and}\quad \frac{dV}{d\lambda}=0\;\bigg(\mathrm{or}\; V'(r_o)=0\bigg)\;.
	\end{align}
	\item [(3).] The energy and angular momentum of the circular orbits are conserved along the geodesic:
	\begin{align}
		V(r_o)=0\quad\mathrm{and}\quad \frac{dE_o}{d\lambda}= \frac{dL_o}{d\lambda}=0\;\bigg(\mathrm{or}\; E_o'(r_o)=L_o'(r_o)=0\bigg)\;.
	\end{align}
\end{itemize}


\subsection{Quasi-Local Definition}
The quasi-local definition of a particle surface provides a coordinate-invariant characterization of circular orbits~\cite{Claudel:2000yi,Song:2022zns}. Here, we review this definition and describe how spacetime circular orbits are obtained.

\dn{ \it {Let $(\mathcal{S},D_a,h_{ab})$ be a timelike hypersurface (or a subset of a timelike hypersurface) of $(\mathcal{M},\nabla_a,g_{ab})$. Let $v^a$ be a unit normal vector to $\mathcal{S}$, satisfying $v^av_a=1$. The metric $g_{ab}$ can be decomposed as
		\begin{align}
			g_{ab}=h_{ab}+v_av_b\;,
		\end{align}
		Let $\gamma$ be a geodesic of a point particle that intersects $\mathcal{S}$ at point $p$. At $p$, the tangent vector $K^a$ of the geodesic can be decomposed as
		\begin{align}
			\label{kdecompose}
			K^a=K_{\parallel}^a+K^a_{\perp}=h^a{}_bK^b+v^av_bK^b=k^a+v^av_bK^b\;.
		\end{align}
		where $k^a\equiv K_{\parallel}^a=h^a{}_bK^b$ is parallel to $\mathcal{S}$ and $K^a_{\perp}=v^av_bK^b$ is normal to $\mathcal{S}$. If for $\forall p\in S$, there exists at least one $\gamma\in S$ passing through $p$ such that
		\begin{align}
			\label{def1}
			K^av_a|_p=0\;,
		\end{align}
		and
		\begin{eqnarray}
			\label{def2}
			K^b\nabla_b (k^ak_{a})|_p=0\;,
		\end{eqnarray}
		then $\mathcal{S}$ is called a (partial) particle surface.
}}

We now apply this definition to derive the evolution equations for circular orbits in a dynamical spherically symmetric spacetime.  For a dynamical spherically symmetric spacetime, the line element in standard spherical coordinates $(t,r,\theta,\phi)$ can be written as
\begin{align}
	\label{dynamicalmetric}
	ds^2=-f(t,r)dt^2+g(t,r)dr^2+r^2(d\theta^2+\sin^2\theta d\phi^2)\;,
\end{align}
where $f$ and $g$ are functions of $t$ and $r$. The unit normal vector $v^a$ to the particle surface in a general dynamical spherical symmetric spacetime is~\cite{Claudel:2000yi}
\begin{align}
	\label{dynamicalv}
	v^a=\frac{\dot{r}\sqrt{g}}{\sqrt{f(f-g\dot{r}^2)}}\bigg(\frac{\partial}{\partial t}\bigg)^a+\frac{\sqrt{f}}{\sqrt{g(f-g\dot{r}^2)}}\bigg(\frac{\partial}{\partial r}\bigg)^a
\end{align}
where a dot denotes derivative with respect to $t$. Due of the spherical symmetry of the system, we again restrict to the equatorial plane, i.e., $\theta=\pi/2$. The tangent vector of a geodesic $\gamma$ is $K^a=\{k^t=\frac{dt}{d\lambda},k^r=\frac{dr}{d\lambda},0,k^\phi=\frac{d\phi}{d\lambda}\}$. The orbital angular momentum $L$ is conserved:
\begin{align}
	\label{Ldynamical}
	L=K_a\bigg(\frac{\partial}{\partial \phi}\bigg)^a=r^2k^\phi=r^2\frac{d\phi}{d\lambda}\;.
\end{align}
From Eq.(\ref{def1}), we obtain the first condition for evolving circular orbits:
\begin{eqnarray}
	\label{dynamicalkv}
	K^av_a|_p=\dot{r}_ok^t-k^r=0\Longrightarrow \frac{dr_o}{d\lambda}=\dot{r}_o\frac{dt}{d\lambda}
\end{eqnarray}
Using Eq.(\ref{Ldynamical}) and condition (\ref{dynamicalkv}), we find $k^a=\{k^t,\dot{r}_ok^t,0,\frac{L_o}{r_o^2}\}$, and
\begin{align}
	k^ak_a=(-f+g\dot{r}^2_o)\bigg(\frac{dt}{d\lambda}\bigg)^2+\frac{L^2}{r_o^2}\;.
\end{align}
From Eq.(\ref{def2}), and defining
\begin{align}
\mathcal{V}=k^ak_a-\epsilon=(-f+g\dot{r}^2_o)\bigg(\frac{dt}{d\lambda}\bigg)^2+\frac{L^2}{r_o^2}-\epsilon\;,
\end{align}
 we have
\begin{align}
	K^a\nabla_a(k^bk_b)|_p=K^a\nabla_a\mathcal{V}|_p=\frac{dt}{d\lambda}\dot{\mathcal{V}}+\frac{dr_o}{d\lambda}\mathcal{V}'=0\;.
\end{align}
Thus, we obtain the second condition for evolving circular orbits:
\begin{align}
\label{dynamicalkV}
	\frac{\dot{\mathcal{V}}}{\dot{r}_o}+\mathcal{V}'=\frac{d\mathcal{V}}{dr_o}=0\;.
\end{align}


\section{Dynamical Potential Formalism}\label{section3}
The quasi-local conditions (\ref{def1}) and (\ref{def2}) are geometrically sound but can be cumbersome for direct computation. We now show how they motivate the introduction of a dynamical potential.

\begin{figure}[H]
	\centering
	\includegraphics[width=2.5in]{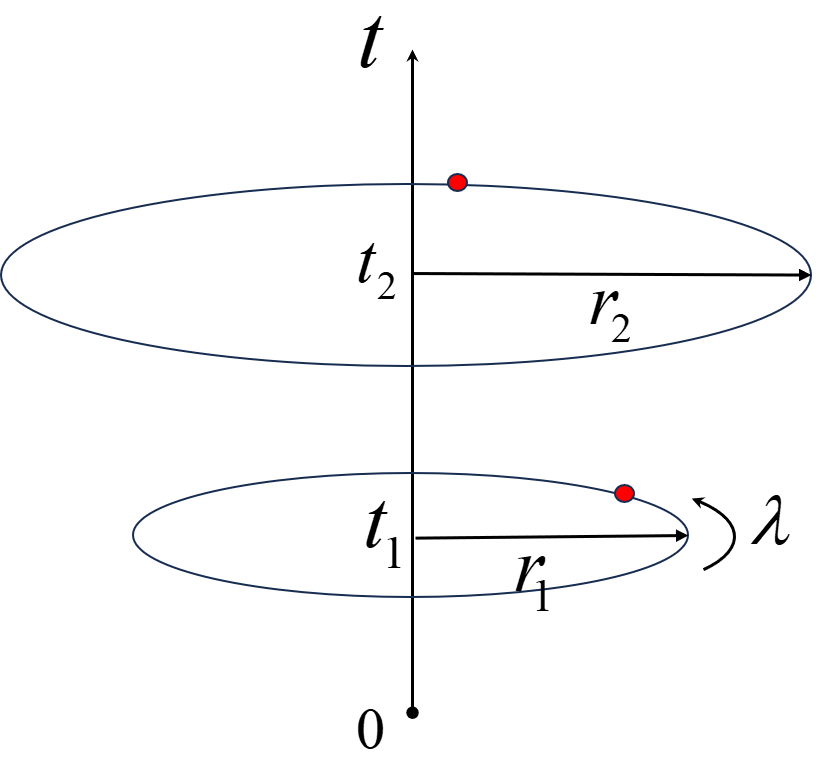}
	\caption{The red point represents a moving point particle. $r_1$
		and $r_2$ are the radii of its two circular orbits at $t=t_1$
		and $t=t_2$, respectively. When $t$ is constant, e.g., $t=t_1$, the radius $r(\lambda,t_1)=r_1$ remains unchanged, and thus does not vary with $\lambda$. This indicates that for an evolving circular orbit in dynamic spacetime, $r$ does not explicitly depend on $\lambda$. }
	\label{fig1}
\end{figure}

The Lagrangian of the system (\ref{dynamicalmetric}) on the equatorial plane is
\begin{align}
\label{LagrangianDynamical}
\mathcal{L}=\frac{1}{2}\bigg[-f(t,r)\bigg(\frac{dt}{d\lambda}\bigg)^2+g(t,r)\bigg(\frac{dr}{d\lambda}\bigg)^2+r^2\bigg(\frac{d\phi}{d\lambda}\bigg)^2\bigg]\;.
\end{align}
From the normalization of the four-velocity $(\frac{dx^\mu}{d\lambda})$, i.e., 
\begin{align}
	\label{normalization2}
	g_{\mu\nu}\frac{dx^\mu}{d\lambda}\frac{dx^\nu}{d\lambda}=\epsilon\;,
\end{align}
we have
\begin{align}
\label{normalize}
-f(t,r)\bigg(\frac{dt}{d\lambda}\bigg)^2+g(t,r)\bigg(\frac{dr}{d\lambda}\bigg)^2+r^2\bigg(\frac{d\phi}{d\lambda}\bigg)^2=\epsilon\;.
\end{align}
Combining Eqs.(\ref{Ldynamical}) and (\ref{normalize}), we obtain
\begin{align}
\label{dr}
	\frac{dr}{d\lambda}=\pm\sqrt{\frac{1}{g(t,r)}\bigg[f(t,r)\bigg(\frac{dt}{d\lambda}\bigg)^2-\frac{L^2}{r^2}+\epsilon\bigg]}\;.
\end{align}
Since $\frac{dt}{d\lambda}$ is a function of $\lambda$, and the right-hand side of Eq.(\ref{dr}) depends on $t$ and $\lambda$, we regard $r$ as $r(\lambda,t(\lambda))$\footnote{In static and stationary spacetime, it reverts to $r(\lambda)$.}. Differentiating $r(\lambda,t(\lambda))$ with respect to the affine parameter gives
\begin{align}
\label{R1}
&\frac{dr}{d\lambda}=\frac{\partial r}{\partial\lambda}+\dot{r}\frac{dt}{d\lambda}\;,\\
\label{R2}
&\frac{d^2r}{d\lambda^2}=\frac{\partial^2r}{\partial\lambda^2}+2\frac{\partial^2 r}{\partial t\partial\lambda}\frac{dt}{d\lambda}+\ddot{r}\bigg(\frac{dt}{d\lambda}\bigg)^2+\dot{r}\frac{d^2t}{d\lambda^2}\;,
\end{align}
Condition (\ref{dynamicalkv}) implies
\begin{align}
	\label{dynamicalr1}
	\frac{\partial r_o}{\partial\lambda}=0\;.
\end{align}
Combining Eqs.(\ref{dr}) and (\ref{R1}), we have
\begin{align}
\frac{\partial r}{\partial\lambda}=\pm\sqrt{\frac{1}{g(t,r)}\bigg[f(t,r)\bigg(\frac{dt}{d\lambda}\bigg)^2-\frac{L^2}{r^2}+\epsilon\bigg]}-\dot{r}\frac{dt}{d\lambda}\;.
\end{align}
We define the dynamical potential $V_{\pm}(t, r)$ as
\begin{align}
	\label{Vtr}
	V_{\pm}(t,r)\equiv\frac{\partial r}{\partial\lambda}=\frac{1}{\sqrt{g}}\bigg[-\sqrt{g}\dot{r}\frac{dt}{d\lambda}\pm\sqrt{f\bigg(\frac{dt}{d\lambda}\bigg)^2+\epsilon-\frac{L^2}{r^2}}\bigg]\;,
\end{align}
The first quasi-local condition (\ref{def1}) or (\ref{dynamicalkv}) is equivalent to
\begin{align}
V_{\pm}(t,r_o)=0\;.
\end{align}
At the location of the evolving circular orbits, we have
\begin{align}
\frac{d V_{\pm}}{d\lambda}=\mp\frac{1}{2\sqrt{g}\sqrt{f(\frac{dt}{d\lambda})^2+\epsilon-\frac{L^2}{r^2}}}\frac{d\mathcal{V}}{d\lambda}\;,
\end{align}
where we have used $V_\pm=0$. Therefore, the second quasi-local condition (\ref{def2}) or (\ref{dynamicalkV}) is equivalent to\footnote{From
\begin{align}
V_{\pm}(t,r)=\frac{\partial r}{\partial\lambda}\;,
\end{align}
and the first condition of the evolving circular orbits, i.e.,
\begin{align}
\frac{\partial r_o}{\partial\lambda}=0\;,
\end{align}
we obtain the condition
\begin{align}
\frac{d V_{\pm}}{d\lambda}=\frac{d}{d\lambda}\bigg(\frac{\partial r_o}{\partial\lambda}\bigg)=\frac{\partial^2r_o}{\partial\lambda^2}+\frac{\partial^2r_o}{\partial t\partial\lambda}\frac{dt}{d\lambda}=0\;.
\end{align}
Thus, we have
\begin{align}
	\frac{d V_{\pm}}{d\lambda}=\frac{\partial^2r_o}{\partial\lambda^2}=0\;.
\end{align}
}
\begin{align}
\frac{d\mathcal{V}}{d\lambda}=0\Longrightarrow \frac{d V_{\pm}}{d\lambda}=0\Longrightarrow\frac{\partial^2r_o}{\partial\lambda^2}=0\;,
\end{align}
Based on the above conclusions, the conditions for an evolving circular orbit are:
\begin{align}
	\label{dynamicalrr1}
	\frac{\partial r_o}{\partial\lambda}=0\Longrightarrow V_{\pm}(t,r_o)=0\;,
\end{align}
and
\begin{align}
\frac{\partial^2r_o}{\partial\lambda^2}=0\Longrightarrow \frac{d V_{\pm}}{d\lambda}=\frac{d\mathcal{V}}{d\lambda}=0\;.
\end{align}
Therefore,
\begin{align}
	\label{dynamicalconditions1}
	&\frac{dr_{o}}{d\lambda}=\dot{r}_o\frac{dt}{d\lambda}\;,\\
	\label{dynamicalconditions2} 
	&\frac{d^2r_o}{d\lambda^2}=\ddot{r}_o\bigg(\frac{dt}{d\lambda}\bigg)^2+\dot{r}_o\frac{d^2t}{d\lambda^2}\;,
\end{align}
The conditions $\frac{\partial r_o}{\partial\lambda}=\frac{\partial^2 r_o}{\partial\lambda^2}=0$ imply that $r$ depends on the affine parameter $\lambda$ only through its time dependence, not directly on $\lambda$, as illustrated in Fig.~\ref{fig1}. The above arguments explain why the evolution equations for circular orbits in dynamical spacetime can be derived from Eqs. (A.1) and (A.5) in~\cite{Mishra:2019trb}, as well as Eqs. (4.1) and (4.8) in~\cite{Song:2021ziq}. Following the steps outlined in~\cite{Mishra:2019trb,Song:2021ziq}, and using Eqs. (\ref{LagrangianDynamical}), (\ref{dynamicalconditions1}), (\ref{dynamicalconditions2}), and the geodesic equations,one can derive the evolution equations for circular orbits in a general dynamical spherically symmetric spacetime.

The dynamical potential $V_{\pm}(t, r)$ is coordinate-dependent and does not arise from a spacetime symmetry. However, it provides a pragmatic and computationally tractable tool for identifying circular orbits in dynamical spacetimes, especially in numerical relativity and astrophysical modeling where a preferred foliation often exists. The equivalence we have established with the quasi-local definition ensures that the resulting orbits are geometrically well-defined and invariant under changes of coordinates away from the chosen slicing. Thus, while the dynamical potential itself is not an invariant quantity, the orbits it selects are.
Alternatively, from the condition $V_\pm=0$, we find that the angular momentum of the circular orbit must satisfy
\begin{align}
L_o^2=r_o^2\bigg\{[f(t,r_o)-g(t,r_o)\dot{r}_o]\bigg(\frac{dt}{d\lambda}\bigg)^2+\epsilon\bigg\}\;.
\end{align}
Note that $L_o$ is conserved along the evolved circular orbits in dynamical spacetime, i.e.,
\begin{align}
\frac{dL_o}{d\lambda}=0\;.
\end{align}
From
\begin{align}
\frac{dL_o}{d\lambda}=\dot{L}_o\frac{dt}{d\lambda}+L_o'\dot{r}_o\frac{dr_o}{d\lambda}=\bigg(\frac{\dot{L}_o}{\dot{r}_o}+L_o'\bigg)\dot{r}_o\frac{dt}{\lambda}=\frac{dL_o}{dr_o}\dot{r}_o\frac{dt}{d\lambda}\;,
\end{align}
we obtain the third form of the equations for evolving circular orbits:
\begin{align}
	V_{\pm}=0\;,\quad\mathrm{and}\quad\frac{dL_o}{d \lambda}=0\;,
\end{align}
or
\begin{align}
	V_{\pm}=0\;,\quad\mathrm{and}\quad\frac{dL_o}{d r_o}=\frac{\dot{L}_o}{\dot{r}_o}+L_o'=0\;.
\end{align}

We now summarize three equivalent conditions for circular orbits in dynamical spacetimes:
\begin{itemize}
	\item [(1).] $r_o$ does not depend explicitly on the affine parameter $\lambda$, i.e.,
	\begin{align}
		\label{1}
		&\frac{\partial r_o}{\partial\lambda}=0\Longrightarrow 	\frac{dr_o}{d\lambda}=\dot{r}_o\frac{dt}{d\lambda}\\
		\label{2}
		&\frac{\partial^2r_o}{\partial\lambda^2}=0\Longrightarrow\frac{d^2r_o}{d\lambda^2}=\ddot{r}_o\bigg(\frac{dt}{d\lambda}\bigg)^2+\dot{r}_o\frac{d^2t}{d\lambda^2}\;.
	\end{align}
	\item [(2).] The dynamical potential of the evolving circular orbits does not change along the geodesic:
	\begin{align}
	\label{3}
		V_{\pm}=0\quad\mathrm{and}\quad \frac{d V_{\pm}}{d \lambda}=\frac{d \mathcal{V}}{d \lambda}=0\;\bigg(\mathrm{or}\;\frac{dV_\pm}{dr_o}= \frac{d\mathcal{V}}{d r_o}=0\bigg)\;.
	\end{align}
	\item [(3).]  If the geodesics in the dynamical spacetime have conserved angular momentum, then for the evolving circular orbit:
	\begin{align}
		\label{4}
		V_{\pm}=0\quad\mathrm{and}\quad \frac{dL_o}{d\lambda}=0\;\bigg(\mathrm{or}\;\frac{dL_o}{dr_o}=0\bigg)\;.
	\end{align}
\end{itemize}
Using the above conditions and the geodesic equations for (\ref{dynamicalmetric}), one can derive the evolution equations for circular orbits in a general dynamical spacetime. These results are exactly consistent with those in~\cite{Song:2022fdg,Song:2021ziq}.

The steps to derive the evolution equations for circular orbits in dynamical spacetimes are as follows: First, regard $r$ as $r(\lambda,t(\lambda))$ to obtain Eqs.(\ref{R1}) and (\ref{R2}), where $r$ is the orbital radius. Next, substitute Eqs.(\ref{R1}) and (\ref{R2}) into Eq. (\ref{normalization2}) to obtain the dynamical potential. Finally, using the above conditions and the geodesic equations, one can derive the evolution equations for evolving circular orbits in general dynamical spacetimes. 

\section{Examples}\label{section4}
We now validate our dynamical potential formalism by applying it to the in-going Vaidya spacetime, the Oppenheimer-Snyder (OS) collapse model and the Lema\^{i}tre-Tolman-Bondi (LTB) model.
\subsection{Vaidya spacetime}
The metric of the 4-dimensional Vaidya spacetime in the in-going null coordinate $\{v,r,\theta,\phi\}$ can be written as~\cite{Vaidya:1951zza}
\begin{eqnarray}
	ds^2=-\bigg[1-\frac{2M(v)}{r}\bigg]dv^2+2dvdr+r^2(d\theta^2+\sin^2\theta d\phi)\;,
\end{eqnarray}
where $M(v)$ is a freely specifiable function of $v$. Due to spherical symmetry, we study the motion of a test particle in the equatorial plane. The Lagrangian is
\begin{align}
\label{Lagrangianvaidya}
\mathcal{L}=\frac{1}{2}\bigg\{-\bigg[1-\frac{2M(v)}{r}\bigg]\bigg(\frac{dv}{d\lambda}\bigg)^2+2\frac{dv}{d\lambda}\frac{dr}{d\lambda}+r^2\bigg(\frac{d\phi}{d\lambda}\bigg)^2\bigg\}\;.
\end{align}
The angular momentum is
\begin{align}
\label{Lvaidya}
L=r^2\frac{d\phi}{d\lambda}\;,
\end{align}
For a null geodesic, the four-velocity $(\frac{dx^\mu}{d\lambda})$  normalization gives
\begin{align}
\label{normalizevaidya}
-\bigg[1-\frac{2M(v)}{r}\bigg]\bigg(\frac{dv}{d\lambda}\bigg)^2+2\frac{dv}{d\lambda}\frac{dr}{d\lambda}+r^2\bigg(\frac{d\phi}{d\lambda}\bigg)^2=0\;,
\end{align}
From the geodesic equation, i.e.,
\begin{align}
\label{geodesic}
	\frac{d^2x^\mu}{d\lambda^2}+\Gamma^\mu{}_{\nu\rho}\frac{dx^\nu}{d\lambda}\frac{dx^\rho}{d\lambda}=0\;,
\end{align}
we have
\begin{align}
\label{vvaidya}
&\frac{d^2v}{d\lambda^2}+\frac{M(v)}{r^2}\bigg(\frac{dv}{d\lambda}\bigg)^2-r\bigg(\frac{d\phi}{d\lambda}\bigg)^2=0\;,\\
\label{rvaidya}
&\frac{d^2r}{d\lambda^2}+\frac{M(v)r+r^2M'(v)-2M(v)^2}{r^3}\bigg(\frac{dv}{d\lambda}\bigg)^2-\frac{2M(v)}{r^2}\frac{dv}{d\lambda}\frac{dr}{d\lambda}+[2M(v)-r]\bigg(\frac{d\phi}{d\lambda}\bigg)^2=0\;.
\end{align}

\begin{itemize}
	\item [(1).] The raidus of the evolving null circular orbits is $r(\lambda,v)$, and the Eqs. (\ref{R1}) and (\ref{R2}) become
	\begin{align}
		\label{VaidyaR1}
		&\frac{dr}{d\lambda}=\frac{\partial r}{\partial\lambda}+\dot{r}\frac{dv}{d\lambda}\;,\\
		\label{VaidyaR2}
		&\frac{d^2r}{d\lambda^2}=\frac{\partial^2r}{\partial\lambda^2}+2\frac{\partial^2 r}{\partial v\partial\lambda}\frac{dv}{d\lambda}+\ddot{r}\bigg(\frac{dv}{d\lambda}\bigg)^2+\dot{r}\frac{d^2v}{d\lambda^2}\;,
	\end{align}
	where a dot denotes a derivative with respect to $v$. Conditions (\ref{1})and (\ref{2}) become
	\begin{align}
		\label{vaidya1}
		&\frac{\partial r_o}{\partial\lambda}=0\Longrightarrow 	\frac{dr_o}{d\lambda}=\dot{r}_o\frac{dv}{d\lambda}\\
		\label{vaidya2}
		&\frac{\partial^2r_o}{\partial\lambda^2}=0\Longrightarrow\frac{d^2r_o}{d\lambda^2}=\ddot{r}_o\bigg(\frac{dv}{d\lambda}\bigg)^2+\dot{r}_o\frac{d^2v}{d\lambda^2}\;.
	\end{align}
	Combining Eqs.(\ref{normalizevaidya}), (\ref{vvaidya}), (\ref{rvaidya}), (\ref{vaidya1}) and (\ref{vaidya2}), we obtain the evolution equation for null circular orbits in Vaidya spacetime:
	\begin{align}
	\label{vaidyar2}
	\ddot{r}_o=\frac{1}{r}\bigg[\bigg(1-\frac{3M(v)}{r_o}\bigg)\bigg(1-\frac{2M(v)}{r_o}-3\dot{r}_o\bigg)-\dot{M}(v)+2\dot{r}_o^2\bigg]\;,
	\end{align}
	Which matches the result in~\cite{Claudel:2000yi,Song:2022fdg}.
	\item [(2).] Substituting Eq.(\ref{VaidyaR1}) into (\ref{normalizevaidya}) and using Eq.(\ref{Lvaidya}), we obtain the dynamical potential in Vaidya spacetime:
	\begin{align}
	V=\frac{1}{2\frac{dv}{d\lambda}}\bigg[\bigg(1-\frac{2M(v)}{r}\bigg)\bigg(\frac{dv}{d\lambda}\bigg)^2-\frac{L^2}{r^2}-2\dot{r}\bigg(\frac{dv}{d\lambda}\bigg)^2\bigg]
	\end{align}
	From condition (\ref{3}) and combining Eqs. (\ref{Lvaidya}), (\ref{normalizevaidya}), and (\ref{vvaidya}), we recover Eq. (\ref{vaidyar2}).
	\item [(3).] From Eqs.(\ref{Lvaidya}) and (\ref{normalizevaidya}), the angular momentum is
    \begin{align}
    \label{vaidyaL2}
    L^2=r^2\bigg[\bigg(1-\frac{2M(v)}{r}\bigg)\bigg(\frac{dv}{d\lambda}\bigg)^2-2\frac{dr}{d\lambda}\frac{dv}{d\lambda}\bigg]\;.
    \end{align}
    From condition (\ref{4}) and combining Eqs. (\ref{vvaidya}), (\ref{vaidya1}), and (\ref{vaidya2}), we again obtain Eq. (\ref{vaidyar2}).
\end{itemize}
For timelike geodesics in the Vaidya spacetime, which satisfy
\begin{align}
	-\bigg[1-\frac{2M(v)}{r}\bigg]\bigg(\frac{dv}{d\lambda}\bigg)^2+2\frac{dv}{d\lambda}\frac{dr}{d\lambda}+r^2\bigg(\frac{d\phi}{d\lambda}\bigg)^2=-1\;,
\end{align}
following the same steps yields the evolved timelike circular orbits satisfying
\begin{eqnarray}
	L_o^2=\frac{2M^2(v)r_o^2+M(v)r_o^3(3\dot{r}_o-1)-r_o^4[\dot{M}(v)+r_o\ddot{r}_o]}{M(v)r_o(5-9\dot{r}_o)-6M^2(v)+r_o^2[r_o\ddot{r}_o+3\dot{r}_o-2\dot{r}_o^2+\dot{M}(v)-1]}\;.
\end{eqnarray}
which is consistent with the result in~\cite{Song:2021ziq,Song:2022fdg}.

\subsection{Oppenheimer-Snyder model}
The OS model describes the dynamical formation of a black hole from homogeneous dust. The metric is that of a FLRW universe ~\cite{Cao:2019vlu,Rezzolla04}
\begin{align}
ds^2=-dt^2+a^2(t)[d\chi^2+\sin^2\chi (d\theta^2+\sin^2\theta d\phi^2)]\;,
\end{align}
where the function $a(t)$ is the scale factor. The Lagrangian for a test particle in the equatorial plane is
\begin{align}
	\label{LagrangianOS}
	\mathcal{L}=\frac{1}{2}\bigg[-\bigg(\frac{dt}{d\lambda}\bigg)^2+a^2\bigg(\frac{d\chi}{d\lambda}\bigg)^2+a^2\sin^2\chi\bigg(\frac{d\phi}{d\lambda}\bigg)^2\bigg]\;.
\end{align}
The angular momentum is
\begin{align}
	\label{LOS}
	L=a^2\sin^2\chi\frac{d\phi}{d\lambda}\;.
\end{align}
For a null geodesic, the four-velocity $(\frac{dx^\mu}{d\lambda})$ normalization gives
\begin{align}
	\label{normalizeOS}
	-\bigg(\frac{dt}{d\lambda}\bigg)^2+a^2\bigg(\frac{d\chi}{d\lambda}\bigg)^2+a^2\sin^2\chi\bigg(\frac{d\phi}{d\lambda}\bigg)^2=0\;.
\end{align}
From the geodesic equation (\ref{geodesic}), we have
\begin{align}
\label{geodesict}
&\frac{d^2t}{d\lambda^2}+a\dot{a}\bigg(\frac{d\chi}{d\lambda}\bigg)^2+a\dot{a}\sin^2\chi\bigg(\frac{d\phi}{d\lambda}\bigg)^2=0\;,\\
\label{geodesicr}
&\frac{d^2\chi}{d\lambda^2}+2\frac{\dot{a}}{a}\frac{dt}{d\lambda}\frac{d\chi}{d\lambda}-\sin\chi\cos\chi \bigg(\frac{d\phi}{d\lambda}\bigg)^2=0\;.
\end{align}
The raidus of the evolving null circular orbits is
\begin{align}
	r(\lambda,t)=a(t)\sin\chi(\lambda,t)\;,
\end{align}
and the Eqs. (\ref{R1}) and (\ref{R2}) reduce to
\begin{align}
	\label{R1OS}
	&\frac{d\chi}{d\lambda}=\frac{\partial \chi}{\partial\lambda}+\dot{\chi}\frac{dt}{d\lambda}\;,\\
	\label{R2OS}
	&\frac{d^2\chi}{d\lambda^2}=\frac{\partial^2\chi}{d\lambda^2}+2\frac{\partial^2 \chi}{\partial t\partial\lambda}\frac{dt}{d\lambda}+\ddot{\chi}\bigg(\frac{dt}{d\lambda}\bigg)^2+\dot{\chi}\frac{d^2t}{d\lambda^2}\;,
\end{align}
where a dot represents the differential with respect to $t$. 
\begin{itemize}
	\item [(1).] Conditions (\ref{1}) and (\ref{2}) become
	\begin{align}
		\label{1OS}
		&\frac{\partial \chi_o}{\partial\lambda}=0\Longrightarrow\frac{d\chi_o}{d\lambda}=\dot{\chi}_o\frac{dt}{d\lambda}\;,\\
		\label{2OS}
		&\frac{\partial^2\chi_o}{d\lambda^2}=0\Longrightarrow\frac{d^2\chi_o}{d\lambda^2}=\ddot{\chi}_o\bigg(\frac{dt}{d\lambda}\bigg)^2+\dot{\chi}_o\frac{d^2t}{d\lambda^2}\;.
	\end{align}
	Combining Eqs.(\ref{normalizeOS}), (\ref{geodesict}), (\ref{geodesicr}), (\ref{1OS}), and (\ref{2OS}), we obtain the evolution equation for null circular orbits in the OS model:
	\begin{align}
	\label{chi2}
		\ddot{\chi}_o=\frac{-a\dot{a}\dot{\chi}_o+\cot\chi_o- a^2\dot{\chi}_o^2\cot\chi_o}{a^2}\;,
	\end{align}
	which matches the result in~\cite{Cao:2019vlu}.
\end{itemize}
\begin{itemize}
	\item [(2).] Substituting Eq.(\ref{R1OS}) into (\ref{normalizeOS}), we obtain the dynamical potential in the OS model:
	\begin{align}
	V_\pm=\frac{1}{a}\bigg[-a\dot{\chi}\frac{dt}{d\lambda}\pm\sqrt{\frac{dt}{d\lambda}-\frac{L^2}{a^2\sin^2\chi}}\bigg]\;,
	\end{align}
where we have used Eq.(\ref{LOS}). From the condition (\ref{3}) and combining Eqs.(\ref{LOS}), (\ref{normalizeOS}), and (\ref{geodesict}), we recover Eq.(\ref{chi2}).

\item [(3).] The angular momentum of the null circular orbits is
\begin{align}
	L_o^2=a^2\sin^2\chi_o(1-a^2\dot{\chi}_o^2)\bigg(\frac{dt}{d\lambda}\bigg)^2\;.
\end{align}
From condition (\ref{4}) and combining Eqs. (\ref{normalizeOS}) and (\ref{geodesict}), we again obtain Eq. (\ref{chi2}).
\end{itemize}
For timelike geodesics in the OS model, which satisfy
\begin{align}
	-\bigg(\frac{dt}{d\lambda}\bigg)^2+a^2\bigg(\frac{d\chi}{d\lambda}\bigg)^2+a^2\sin^2\chi\bigg(\frac{d\phi}{d\lambda}\bigg)^2=-1\;,
\end{align}
following the same steps yields the evolved timelike circular orbits satisfying
\begin{align}
L_o^2=\frac{a^3\sin^2\chi_o[\dot{a}\dot\chi_o(a^2\dot{\chi}_o^2-2)-a\ddot{\chi_o}]}{a(\dot{a}\dot{\chi}_o+a\ddot{\chi}_o)+(a^2\dot{\chi}_o^2-1)\cot\chi_o}\;.
\end{align}

\subsection{Lema\^{i}tre-Tolman-Bondi (LTB) model}
The metric of the Lema\^{i}tre-Tolman-Bondi (LTB) model can be expressed as~\cite{Cao:2019vlu}
\begin{equation}
	\label{LTB}
	ds^2=-dt^2 + \frac{[r_x(t,x)]^2}{1+\kappa(x)}dx^2 + r^2(t,x)(d\theta^2+\sin^2\theta d\phi^2)\;,
\end{equation}
where $\kappa(x)$ is the so-called specific bending energy which is a function of $x$, and $r_x\equiv\partial r/\partial x$. The Lagrangian for a test particle in the equatorial plane is
\begin{align}
	\label{LagrangianLTB}
	\mathcal{L}=\frac{1}{2}\bigg[-\bigg(\frac{dt}{d\lambda}\bigg)^2+\frac{[r_x(t,x)]^2}{1+\kappa(x)}\bigg(\frac{dx}{d\lambda}\bigg)^2+r^2(t,x)\bigg(\frac{d\phi}{d\lambda}\bigg)^2\bigg]\;.
\end{align}
The angular momentum is
\begin{align}
	\label{LLTB}
	L=r^2\frac{d\phi}{d\lambda}\;.
\end{align}
For a null geodesic, the four-velocity $(\frac{dx^\mu}{d\lambda})$  normalization gives
\begin{align}
	\label{normalizeLTB}
	-\bigg(\frac{dt}{d\lambda}\bigg)^2+\frac{[r_x(t,x)]^2}{1+\kappa(x)}\bigg(\frac{dx}{d\lambda}\bigg)^2+r^2(t,x)\bigg(\frac{d\phi}{d\lambda}\bigg)^2=0\;.
\end{align}
From the geodesic equation (\ref{geodesic}), we have
\begin{align}
\label{geodesictLTB}
&\frac{d^2t}{d\lambda^2}+\frac{r_xr_{tx}}{1+\kappa}\bigg(\frac{dx}{d\lambda}\bigg)^2+rr_t\bigg(\frac{d\phi}{d\lambda}\bigg)^2=0\;,\\
\label{geodesicrLTB}
&\frac{d^2x}{d\lambda^2}+2\frac{r_{tx}}{r_x}\frac{dt}{d\lambda}\frac{dx}{d\lambda}+\bigg(\frac{r_{xx}}{r_x}-\frac{k'}{2(1+k)}\bigg)\bigg(\frac{dx}{d\lambda}\bigg)^2-\frac{(1+\kappa)r}{r_x}\bigg(\frac{d\phi}{d\lambda}\bigg)^2=0\;.
\end{align}
where $r_{tx}\equiv\partial^2 r/\partial t\partial x$, $r_{xx}\equiv\partial^2 r/\partial x^2$, $r_{t}\equiv\partial r/\partial t$ and $\kappa'\equiv d\kappa/dx$. The raidus of the evolving null circular orbits is $r(t,x(t,\lambda))$, and the Eqs. (\ref{R1}) and (\ref{R2}) reduce to
\begin{align}
	\label{R1LTB}
	&\frac{dx}{d\lambda}=\frac{\partial x}{\partial\lambda}+\dot{x}\frac{dt}{d\lambda}\;,\\
	\label{R2LTB}
	&\frac{d^2x}{d\lambda^2}=\frac{\partial^2x}{d\lambda^2}+2\frac{\partial^2 x}{\partial t\partial\lambda}\frac{dt}{d\lambda}+\ddot{x}\bigg(\frac{dt}{d\lambda}\bigg)^2+\dot{x}\frac{d^2t}{d\lambda^2}\;,
\end{align}
\begin{itemize}
	\item [(1).] Conditions (\ref{1}) and (\ref{2}) become
	\begin{align}
		\label{1LTB}
		&\frac{\partial x_o}{\partial\lambda}=0\Longrightarrow\frac{dx_o}{d\lambda}=\dot{x}_o\frac{dt}{d\lambda}\;,\\
		\label{2LTB}
		&\frac{\partial^2x_o}{d\lambda^2}=0\Longrightarrow\frac{d^2x_o}{d\lambda^2}=\ddot{x}_o\bigg(\frac{dt}{d\lambda}\bigg)^2+\dot{x}_o\frac{d^2t}{d\lambda^2}\;.
	\end{align}
	Combining Eqs.(\ref{normalizeLTB}), (\ref{geodesictLTB}), (\ref{geodesicrLTB}), (\ref{1LTB}), and (\ref{2LTB}), we obtain the evolution equation for null circular orbits in the LTB model:
	\begin{eqnarray}
		\label{x2LTB}
		\ddot{x}_o=\frac{1+\kappa}{rr_x} + \Big(\frac{r_t}{r}-\frac{2r_{tx}}{r_x}\Big)\dot{x}_o -\Big( \frac{r_x}{r}+\frac{r_{xx}}{r_x}-\frac{1}{2}\frac{\kappa_x}{1+\kappa}\Big)\dot{x}_o^2 + \frac{r_x^2}{1+\kappa}\Big(\frac{r_{xt}}{r_x}-\frac{r_t}{r}\Big)\dot{x}_o^3\, ,
	\end{eqnarray}
	which matches the result in~\cite{Cao:2019vlu}.
	\item [(2).] Substituting Eq.(\ref{R1LTB}) into (\ref{normalizeLTB}) and using Eq.(\ref{LLTB}), we obtain the dynamical potential in the LTB model:
	\begin{align}
		V_\pm=-\dot{x}\frac{dt}{d\lambda}\pm\sqrt{\frac{1+\kappa}{r_x^2}\bigg(\frac{dt}{d\lambda}\bigg)^2-\frac{L^2}{r^2}}\;,
	\end{align}
	From condition (\ref{3}) and combining Eqs. (\ref{LLTB}), (\ref{normalizeLTB}), and (\ref{geodesictLTB}), we recover Eq. (\ref{x2LTB}).
	\item [(3).] From Eqs.(\ref{LLTB}) and (\ref{normalizeLTB}), the angular momentum is
	\begin{align}
		L^2=\frac{1}{r^2}\bigg[\bigg(\frac{dt}{d\lambda}\bigg)^2-\frac{r_x^2}{1+\kappa}\bigg(\frac{dx}{d\lambda}\bigg)^2\bigg]\;.
	\end{align}
	From condition (\ref{4}) and using Eqs. (\ref{normalizeLTB}) and (\ref{geodesictLTB}), we again obtain Eq. (\ref{x2LTB}).
	\end{itemize}
	For timelike geodesics in the LTB model, which satisfy
	\begin{align}
	-\bigg(\frac{dt}{d\lambda}\bigg)^2+\frac{[r_x(t,x)]^2}{1+\kappa(x)}\bigg(\frac{dx}{d\lambda}\bigg)^2+r^2(t,x)\bigg(\frac{d\phi}{d\lambda}\bigg)^2=-1\;,
	\end{align}
	following the same steps yields the evolved timelike circular orbits satisfying
	\begin{align}
	L_o^2=\frac{B}{A}\;,
	\end{align}
	where
	\begin{align}
	B=-r^3[\kappa'r_x\dot{x}_o^2+2r_{tx}\dot{x}_o(r_x^2\dot{x}_o^2-2-2\kappa)-2(1+\kappa)(r_{xx}\dot{x}_o^2+r_x\ddot{x}_o)]\;,
	\end{align}
	and
	\begin{align}
	A&=2\kappa^2-2(1+r_tr_x\dot{x}_o)(r_x^2\dot{x}_o^2-1)+r[(\kappa r_x-\frac{4r_{tx}}{\dot{x}_o}-2r_{xx}+2r_{tx}r_x^2)\dot{x}_o^2-2r_x\ddot{x}_o]&\nonumber\\
	&-2\kappa\{(r_x^2-\frac{r_tr_x}{\dot{x}_o})\dot{x}_o^2-2+r[(\frac{2r_{tx}}{\dot{x}_o}+r_{xx})\dot{x}_o^2+r_x\ddot{x}_o]\}\;.
	\end{align}
\section{Conclusions and discussions}\label{conclusion}
In this work, we have established a unified framework for defining and locating evolving circular orbits in dynamical spherically symmetric spacetimes, bridging the gap between geometric invariance and practical computation. By starting from the quasi-local definition of a particle surface—a coordinate-invariant formulation—we derived a set of conditions that naturally lead to the introduction of a dynamical potential $V_{\pm}(t, r)$. This potential, though coordinate-dependent, provides an intuitive and computationally tractable tool for identifying circular orbits in time-dependent gravitational fields.

We rigorously demonstrated the equivalence between the quasi-local approach and the dynamical potential formalism, ensuring that the orbits identified are geometrically meaningful and independent of the choice of foliation away from the selected coordinates. Moreover, we showed that the condition for circular orbits can also be equivalently expressed in terms of the conservation of angular momentum $L_o$ along the evolving orbit, i.e., $\frac{dL_o}{d\lambda}$, which extends the familiar notion of conserved quantities from static to dynamical settings. This provides a physically intuitive and consistent generalization of the effective potential method.

The consistency and applicability of our method were explicitly verified in the Vaidya spacetime, the Oppenheimer-Snyder (OS) collapse model, and the Lema\^{i}tre-Tolman-Bondi (LTB) model. We successfully recovered the known evolution equations for both null and timelike circular orbits using all three equivalent formulations: via the dynamical potential, the quasi-local geometric conditions, and the angular momentum conservation criterion.

Our results show that the dynamical potential approach not only retains the conceptual clarity of the traditional effective potential method—known for its utility in static and stationary spacetimes—but also extends its applicability to fully dynamical contexts. The alternative formulation using angular momentum further enriches the framework, offering multiple complementary perspectives for analyzing orbital dynamics. This opens up new avenues for studying phenomena such as gravitational collapse, black hole formation, and accretion processes in time-dependent backgrounds.

Future work may involve applying this framework to more complex dynamical spacetimes, such as those with matter accretion or radiation, and exploring its implications for gravitational wave astronomy and numerical relativity. The adaptability of the dynamical potential method, together with the angular momentum-based formulation, makes it a promising tool for both theoretical studies and numerical simulations in evolving gravitational environments.

\section*{Acknowledgement}

We are grateful to Professor Li-Ming Cao for his valuable discussions and kind assistance.



\begin{thebibliography}{99}
\bibitem{EventHorizonTelescope:2019dse}
K.~Akiyama \textit{et al.} [Event Horizon Telescope],
Astrophys. J. Lett. \textbf{875} (2019), L1
doi:10.3847/2041-8213/ab0ec7
[arXiv:1906.11238 [astro-ph.GA]].

\bibitem{Abramowicz:2011xu}
M.~A.~Abramowicz and P.~C.~Fragile,
Living Rev. Rel. \textbf{16} (2013), 1
doi:10.12942/lrr-2013-1
[arXiv:1104.5499 [astro-ph.HE]].

\bibitem{Grenzebach:2014fha}
A.~Grenzebach, V.~Perlick and C.~L\"ammerzahl,
Phys. Rev. D \textbf{89} (2014) no.12, 124004
doi:10.1103/PhysRevD.89.124004
[arXiv:1403.5234 [gr-qc]].

\bibitem{Maggiore:2007ulw}
M.~Maggiore,
Oxford University Press, 2007,
ISBN 978-0-19-171766-6, 978-0-19-852074-0
doi:10.1093/acprof:oso/9780198570745.001.0001


\bibitem{Berry:2020ntz}
T.~Berry, A.~Simpson and M.~Visser,
Universe \textbf{7} (2020) no.1, 2
doi:10.3390/universe7010002
[arXiv:2008.13308 [gr-qc]].

\bibitem{Teo:2003ltt}
E.~Teo,
Gen. Rel. Grav. \textbf{35} (2003) no.11, 1909-1926
doi:10.1023/A:1026286607562


\bibitem{Cardoso:2008bp}
V.~Cardoso, A.~S.~Miranda, E.~Berti, H.~Witek and V.~T.~Zanchin,
Phys. Rev. D \textbf{79} (2009) no.6, 064016
doi:10.1103/PhysRevD.79.064016
[arXiv:0812.1806 [hep-th]].

\bibitem{Teo:2020sey}
E.~Teo,
Gen. Rel. Grav. \textbf{53} (2021) no.1, 10
doi:10.1007/s10714-020-02782-z
[arXiv:2007.04022 [gr-qc]].


\bibitem{Perlick:2021aok}
V.~Perlick and O.~Y.~Tsupko,
Phys. Rept. \textbf{947} (2022), 1-39
doi:10.1016/j.physrep.2021.10.004
[arXiv:2105.07101 [gr-qc]].
	
\bibitem{Claudel:2000yi}
C.~M.~Claudel, K.~S.~Virbhadra and G.~F.~R.~Ellis,
J. Math. Phys. \textbf{42} (2001), 818-838
doi:10.1063/1.1308507
[arXiv:gr-qc/0005050 [gr-qc]].

\bibitem{Cao:2019vlu}
L.~M.~Cao and Y.~Song,
Eur. Phys. J. C \textbf{81} (2021) no.8, 714
doi:10.1140/epjc/s10052-021-09502-0
[arXiv:1910.13758 [gr-qc]].


\bibitem{Kobialko:2022uzj}
K.~Kobialko, I.~Bogush and D.~Gal'tsov,
Phys. Rev. D \textbf{106} (2022) no.8, 084032
doi:10.1103/PhysRevD.106.084032
[arXiv:2208.02690 [gr-qc]].

\bibitem{Qiao:2022jlu}
C.~K.~Qiao and M.~Li,
Phys. Rev. D \textbf{106} (2022) no.2, L021501
doi:10.1103/PhysRevD.106.L021501
[arXiv:2204.07297 [gr-qc]].

\bibitem{Song:2022zns}
Y.~Song, Y.~Cen, L.~Tang, J.~Hu, K.~Diao, X.~Zhao and S.~Shi,
Eur. Phys. J. C \textbf{83} (2023) no.9, 833
doi:10.1140/epjc/s10052-023-11970-5
[arXiv:2208.03665 [gr-qc]].

\bibitem{Song:2022fdg}
Y.~Song and C.~Zhang,
Eur. Phys. J. C \textbf{83} (2023) no.1, 50
doi:10.1140/epjc/s10052-022-11143-w
[arXiv:2208.03661 [gr-qc]].


\bibitem{Bogush:2023ojz}
I.~Bogush, K.~Kobialko and D.~Gal'tsov,
Phys. Rev. D \textbf{108} (2023) no.4, 044070
doi:10.1103/PhysRevD.108.044070


\bibitem{Jefremov:2015gza}
P.~I.~Jefremov, O.~Y.~Tsupko and G.~S.~Bisnovatyi-Kogan,
Phys. Rev. D \textbf{91}, no.12, 124030 (2015)
doi:10.1103/PhysRevD.91.124030
[arXiv:1503.07060 [gr-qc]].

\bibitem{Toshmatov:2020wky}
B.~Toshmatov, O.~Rahimov, B.~Ahmedov and D.~Malafarina,
Eur. Phys. J. C \textbf{80}, no.7, 675 (2020)
doi:10.1140/epjc/s10052-020-8254-6
[arXiv:2003.09227 [gr-qc]].



\bibitem{Mishra:2019trb}
A.~K.~Mishra, S.~Chakraborty and S.~Sarkar,
Phys. Rev. D \textbf{99}, no.10, 104080 (2019)
doi:10.1103/PhysRevD.99.104080
[arXiv:1903.06376 [gr-qc]].

\bibitem{Song:2021ziq}
Y.~Song,
Eur. Phys. J. C \textbf{81}, no.10, 875 (2021)
doi:10.1140/epjc/s10052-021-09623-6
[arXiv:2108.00696 [gr-qc]].


\bibitem{Vaidya:1951zza}
P.~C.~Vaidya,
Phys. Rev. \textbf{83}, 10-17 (1951)
doi:10.1103/PhysRev.83.10

\bibitem{Rezzolla04}
L.~Rezzolla,
``An introduction to gravitational collapse to black holes,''
Lecture notes for International School of Gravitation and Cosmology, (2004).
\end{thebibliography}
\end{document}